\documentclass[12pt,preprint]{aastex}
\usepackage{apjfonts}
\usepackage{graphicx}
\usepackage{lineno}

\usepackage{ulem}
\usepackage{appendix}
\usepackage{xspace}
\usepackage{bm}
\usepackage{color}
\usepackage{CJKutf8}
\usepackage{float}
\usepackage{CJKnumb}

\def\tess{{\it TESS}\xspace}
\def\kepler{{\it Kepler}\xspace}

\begin{document}
\begin{CJK}{UTF8}{gbsn}

\title{KIC 9845907: A $\delta$ Scuti star with the first overtone as the dominant frequency and with  many equidistant structures in its spectrum}

\author{
Xiao-Ya Sun\altaffilmark{1},
Zhao-Yu Zuo\altaffilmark{1, *},
Tao-Zhi Yang\altaffilmark{1},
Antonio Garc\'{\i}a Hern\'{a}ndez\altaffilmark{2}
}

\altaffiltext{1}{
Ministry of Education Key Laboratory for Nonequilibrium Synthesis and Modulation of Condensed Matter, 
School of Physics, Xi'an Jiaotong University, Xi'an 710049, China; *Email: zuozyu@xjtu.edu.cn (ZYZ)}

\altaffiltext{2}{
Departamento de F\'{\i}sica Te{\'orica} y del Cosmos, Universidad de Granada, Campus de Fuentenueva s/n, E-18071, Granada, Spain}


\begin{abstract}

In this paper, we present an analysis of the pulsating behavior of \kepler target KIC 9845907. Using the data from \kepler, we detected 85 significant frequencies, including the first overtone $f_{1}$ = 17.597 day$^{-1}$ as the dominant frequency, the non-radial independent frequency $f_{3}$ = 31.428 day$^{-1}$ ($\ell$=1), as well as two modulation terms $f_{m1}$ = 0.065 day$^{-1}$ and $f_{m2}$ = 1.693 day$^{-1}$. We found fourteen pairs of triplet structures with $f_{m1}$ or $f_{m2}$, four pairs of which can further form quintuplet structures. We note these are the most intriguing features discovered in this study and they were recognized for the first time in $\delta$ Scuti stars. We discussed several possible explanations, i.e., beating, the Blazhko effect, combination mode hypothesis, nonlinear mode coupling, large separation, and stellar rotational splitting for these equidistant structures. Our asteroseismic models indicate this modulation with $f_{m1}$ might be related to the rotational splitting. The study of more $\delta$ Scuti stars with triplet and/or quintuplet structures using high-precision space photometry would be helpful to further explore its origin.

\end{abstract}

\keywords{stars: individual (KIC 9845907) - stars: oscillations - stars: variable: $\delta$ Scuti}

\section{Introduction}

One of the long-standing goals in astronomy is to improve our understanding of stellar internal structure and evolution. Asteroseismology, by comparing observations with theoretical models, constrains the interior physics of stars, such as convection, rotation, etc \citep{Aerts2021}. In recent years, asteroseismology has entered its golden age thanks to the continuous, long-term and high-resolution data from space telescopes such as \kepler \citep{Koch2010} and Transiting Exoplanet Survey Satellite (\tess; \citealt{Ricker2015}). Significant progress has been made in the study of various types of pulsators, such as red giant stars \citep{Yu2020,Li2022}, solar-type stars \citep{Chaplin2013}, pulsating white dwarfs \citep{Zong2016}, and $\delta$ Scuti stars \citep{Chen2019,Yang2022} as well. Among the more than 150,000 observed targets in the \kepler field, over 2000 $\delta$ Scuti stars have been discovered \citep{Balona2014,Bowman2016}, which provides a large number of samples for asteroseismic studies.

$\delta$ Scuti stars are a class of short-period variables with periods in range of 0.02 $-$ 0.25 days. They are intermediate-mass stars and have spectral types from A to F \citep{Breger2000}. Their masses are generally between $1.5 {M_\odot}$ and $2.5 {M_\odot}$, which place them in the transition region between the lower-mass stars with thick convective envelope and radiative core and the massive ones with thin or none convective shell but a convective core \citep{Aerts2010}. They are located in the classical Cepheid instability strip where it crosses the main sequence (MS) in the Hertzsprung$-$Russell (H$-$R) diagram. Most $\delta$ Scuti stars are on the MS or post-MS stages although there are also some on the pre-MS stage \citep{Breger2000,Zwintz2008}. These variables exhibit both radial and non-radial modes \citep{Qian2018,Sun2021}, excited mainly by the $\kappa$ mechanism \citep{Aerts2010}, usually identified as low-radial-order ($n$) low-angular-degree ($l$) pressure ($p$) modes \citep{Uytterhoeven2011,Holdsworth2014}. Moreover, \cite{Uytterhoeven2011} discovered that many of the $\delta$ Scuti stars show also $g-$modes. These characteristics make them excellent targets for asteroseismology study. 

The pulsation spectra of $\delta$ Scuti stars are generally very rich and messy, which challenges the mode identification \citep{Goupil2005,Bedding2023}. Recently, there have been many studies on complex spectra from different perspectives, such as the low-order large separation \citep{2020MNRAS.498.1700R}, rotational splitting \citep{Ram2021}, (near-)equidistant frequency spacing structures (such as triplets and/or quintuplets; \citealt{Kolenberg2011}), and the relation between the low-order large separation and the stellar mean density \citep{2014A&A...563A...7S,2015ApJ...811L..29G}. For instance, in KIC 5950759, two pairs of triplet structures centered on the dominant frequency were detected in its frequency spectra, and the cause of that is inferred to be the amplitude modulation of stellar rotation \citep{Yang2018}. Another $\delta$ Scuti star KIC 10284901 shows two pairs of quintuplet structures, which might be related to the Blazhko effect \citep{Yang2019}. Moreover, \cite{Chen2017} analyzed the frequency spectrum of CoRoT 102749568 based on the rotational splitting of the oscillation mode and determined the stellar parameters and helium core size using asteroseismology. These multiplet structures and the modes identification might improve our knowledge of the $\delta$ Scuti stars and offer new clues for probing the stellar interior and physical processes.

KIC 9845907 was classified as a $\delta$ Scuti star by \cite{Uytterhoeven2011}. The \kepler magnitude of this star is $K_{P}$ = 11.64 mag, and its effective temperature and radius are $T_{\rm eff}$  = 7936 $\pm$ 200 K and $R$ = 1.954 $R_{\odot}$, respectively \citep{Brown2011}. Based on the data of short-cadence (SC) observations in the \kepler field, \cite{Balona2016} found a large number of combination frequencies in $\delta$ Scuti stars, with the number in KIC 9845907 being 10. Some basic parameters of KIC 9845907 are listed in Table \ref{tab:basic_parmeters}. In this work, we used the high-precision photometric data (including SC and LC data) provided by \kepler to further study the pulsating behavior of KIC 9845907.

\begin{deluxetable}{lcc}
\tabletypesize{\small} 
\tablewidth{0pc} 
\tablecaption{Basic Properties of KIC 9845907 
\label{tab:basic_parmeters}} 
\tablehead{ 
\colhead{Parameters}   & 
\colhead{KIC 9845907}      &
\colhead{References}       
}
\startdata 
   $K_{P}$              &  11.64 mag  &   a  \\
   $TESS$ magnitude       &  11.41 mag  &   c  \\
   $B$                  &  11.81 mag  &   b  \\
   $V$                  &  11.12 mag  &   b  \\
   $J$                  &  11.03 mag  &   b  \\
   $I$                  &  11.23 mag  &   b  \\
   $K$                  &  10.94 mag  &   b  \\
   $H$                  &  10.98 mag  &   b  \\
   $g$                  &  11.67 mag  &   b  \\
   $i$                  &  11.72 mag  &   b  \\
   $z$                  &  11.79 mag  &   b  \\
   $Gaia$               &  11.57 mag &    c  \\
   $M/M_{\odot}$        &  1.87 ± 0.29 &  c   \\
   $R/R_{\odot}$         &  1.954  &       b  \\   
                         & 1.928 ± 0.067 &  c  \\ 
   $T_{\rm eff}$            &   7936 ± 200 K    &   a  \\    
                           &   7861 ± 147 K    &   c  \\      
   log $g$                &  4.029 ± 0.5 dex  &   b  \\
                          &  4.140 ± 0.077 dex  &   c  \\  
   ${\rm[Fe/H]}$               &  $-$0.039 ± 0.5 dex &      b  \\
   Parallax (mas)       & 1.25 ± 0.01 &  ($Gaia$),d \\
   \enddata 
   
   \tablecomments{(a) \cite{Balona2016}; (b) KASOC: https://kasoc.phys.au.dk/; (c) TASOC: https://tasoc.dk; (d) $Gaia$ \citep{McDonald2017}.}
\end{deluxetable}

\section{OBSERVATIONS AND DATA REDUCTION}

\begin{figure*}[h]
\begin{center}
  \includegraphics[width=0.95\textwidth,trim=45 0 55 20,clip]{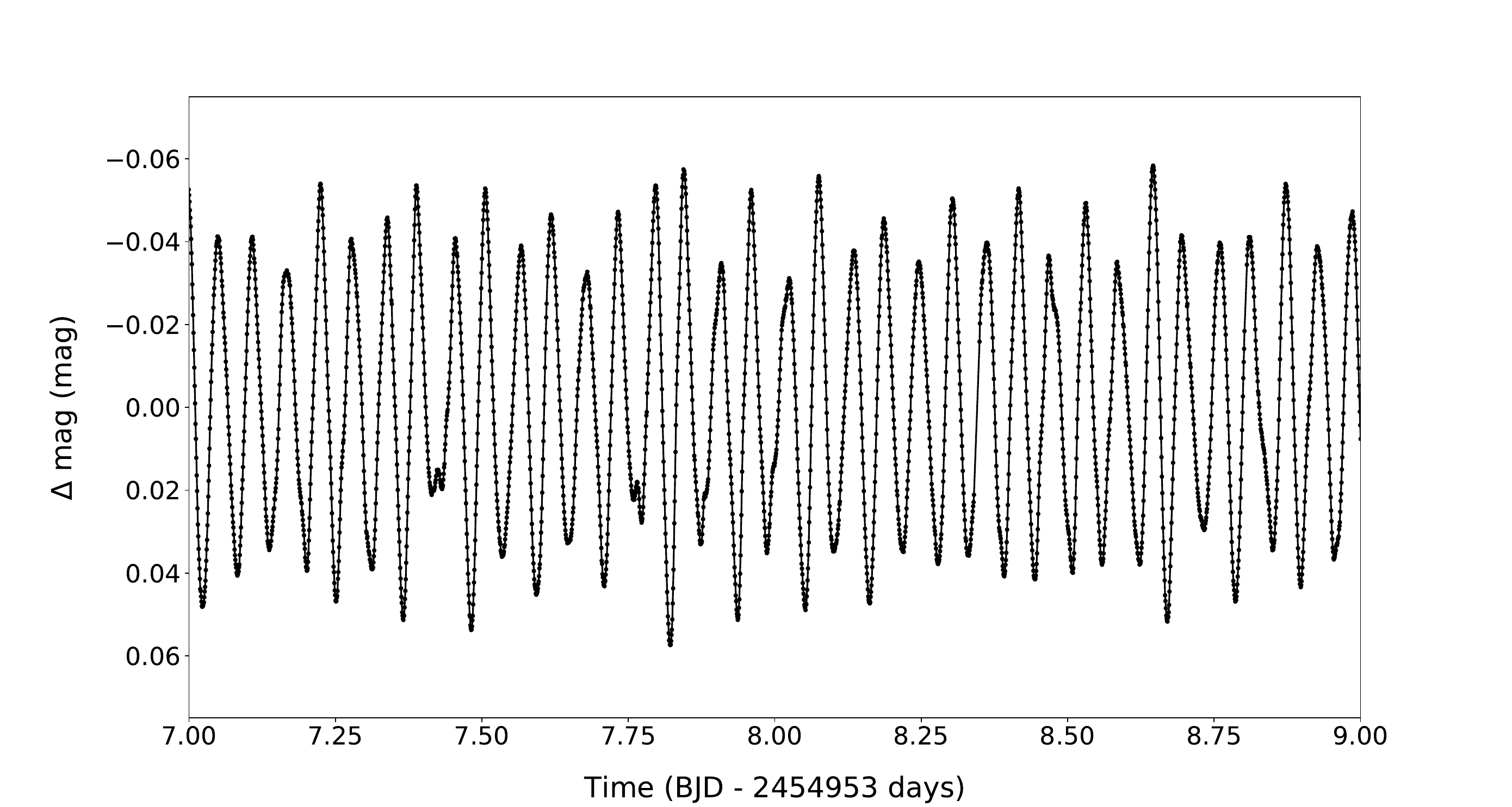}
  \caption{A 2 day section of the \kepler SC light curve of KIC 9845907. The amplitude of the light curve is about 0.1 mag. BJD is barycentric Julian date.}
    \label{fig:light curve}
\end{center}
\end{figure*}

KIC 9845907 was observed by \kepler from BJD 2,454,953.539 to 2,456,424.001 (Q0$-$Q6, Q8$-$Q10, Q12$-$Q14, and Q16$-$Q17, total of 15 quarters) in 29.5 minute cadence (i.e., LC mode) and from BJD 2,454,953.529 to 2,455,833.279 (Q0, Q5.1$-$Q5.3, Q8.1$-$Q8.3, Q9.1$-$Q9.3 and Q10.1$-$Q10.3, total of five quarters) in 58.5 s cadence (i.e., SC mode). All the data were downloaded from \kepler Asteroseismic Science Operations Center (KASOC) database\footnote{KASOC database: {https://kasoc.phys.au.dk}}. KASOC provides asteroseismological data from the NASA \kepler and K2 missions to astronomers. The KASOC archive classifies the pulsation stars into subcategories and $\delta$ Scuti stars corrected by Working Group 4. And KASOC provides two types data: the "raw" and corrected flux data. In this work, we converted the corrected flux to stellar magnitude and performed corrections, including eliminating the outliers and detrending the light curves. Then the mean value of each quarter was subtracted, and the rectified time series were obtained with 49551 data points in LC and 176899 in SC, spanning over about 1471 and 880 days, respectively. Fig.~\ref{fig:light curve} shows a section of the SC light curves of KIC 9845907, where the amplitude is about 0.1 mag.

\section{FREQUENCY ANALYSIS}

In this work, the \kepler SC data of KIC 9845907 were analyzed to search for significant frequencies using the software PERIOD04 \citep{Lenz2005}, which allows a pre-whitening procedure spotting the frequencies through a Fourier transform of the data. The rectified light curve was fitted with the following formula:

\begin{equation}
m = m_{0} + \sum A_{i} {\rm sin}(2\pi(f_{i}t + \phi_{i}))  \label{equation1}
\end{equation}

where $m_{0}$ is the zero point and $A_{i}$, $f_{i}$, $\phi_{i}$ are the amplitude, frequency, and phase of each mode, respectively. We chose a range of 0 $<$ $\nu$ $<$ 100 day$^{-1}$ to search for frequencies since it covers the pulsation regime of $\delta$ Scuti stars. The upper limit of this range is well below the Nyquist frequency ($f_{Ny}=1/(2 \Delta t$) = 734 day$^{-1}$, where $\Delta t$ is the sampling interval between consecutive points). The criterion of a signal-to-noise ratio (S/N) > 4.0 was adopted as stopping criterion for the pre-whitening process \citep{Breger1993}. The noise for each significant frequency was calculated in a box size of 2 day$^{-1}$ centered in the extracted peak. We selected a step rate "high," corresponding to an oversampling of 20 and a frequency spacing of 0.00005683 day$^{-1}$. The uncertainties of all frequencies were calculated following the method provided by \cite{Montgomery1999}. 

A total of 85 significant frequencies with S/N > 4.0 were extracted in this work. A full list of the detected frequencies (i.e., $f_{1}$ to $f_{85}$), with their corresponding amplitudes and S/N, is given in Table~A1 in the Appendix. Fig.~\ref{fig:fre_amp} shows the Fourier spectra of 85 significant frequencies in two different ways: one is the amplitude spectrum of KIC 9845907 in the 0 $-$ 100 day$^{-1}$ range (upper panel), and the other is the distribution of detected 85 frequencies (lower panel). As can be seen clearly in Fig.~\ref{fig:fre_amp}, the frequencies with higher amplitudes were found in the range of 15 $-$ 35 day$^{-1}$, and the other smaller peaks (i.e., $f_{8}$ and $f_{9}$) were in the range of 0 $-$ 2 day$^{-1}$.

\begin{figure*}[h]
\begin{center}
  \includegraphics[width=0.95\textwidth,trim=5 10 20 0,clip]{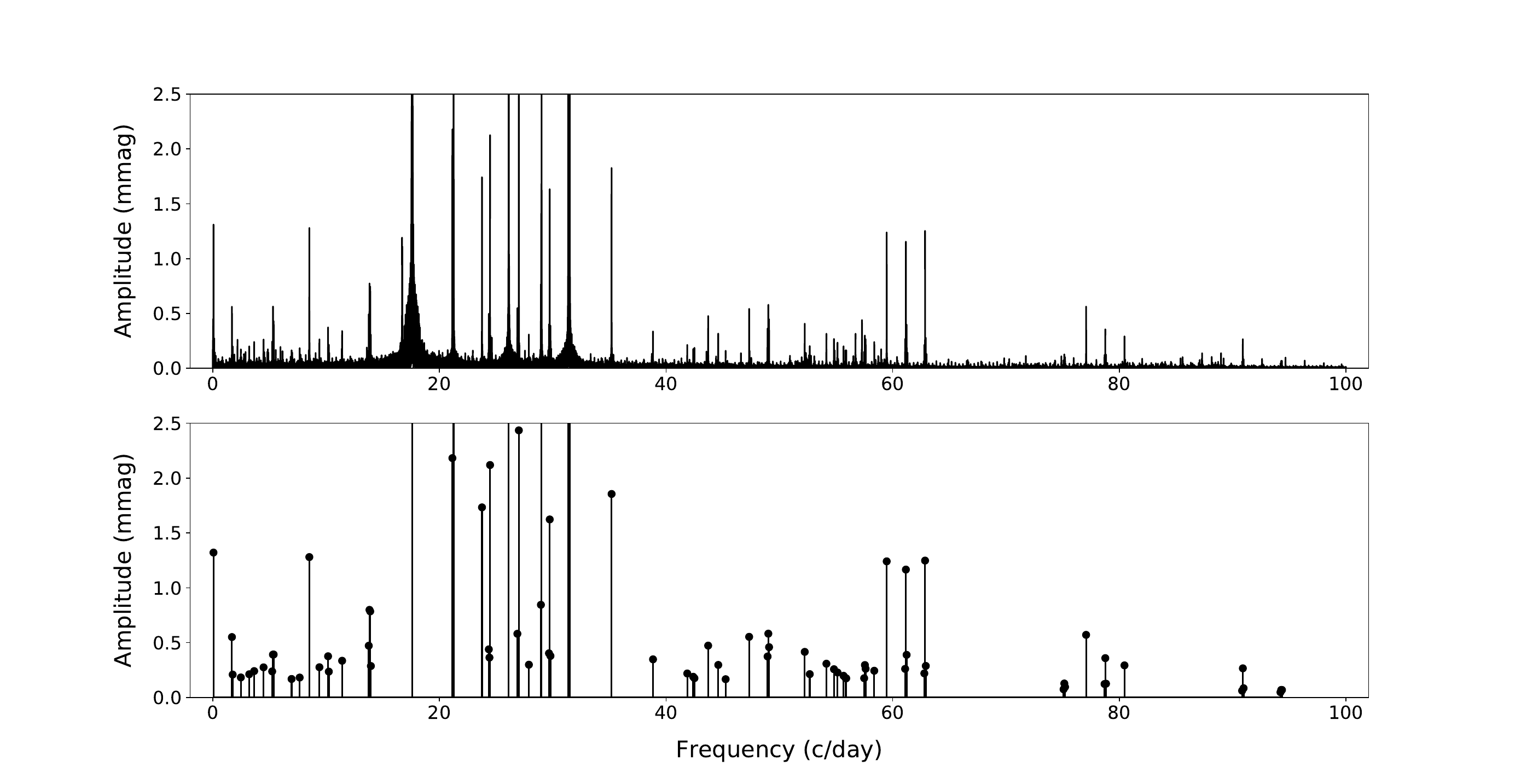}
  \caption{Amplitude spectra of KIC 9845907 using the SC data. The upper panel shows the amplitude spectrum in the 0 $-$ 100 c/day range. The lower panel represents the distribution of detected 85 frequencies.}
    \label{fig:fre_amp}
\end{center}
\end{figure*}

The $\delta$ Scuti stars usually pulsate in radial mode, and theoretically the radial modes have higher visibility than non-radial \citep{Dziembowski1977}. Therefore, by visually inspecting the highest amplitude of the frequency spectra, we considered $f_{1}$ as a radial mode. Further evidence comes from the investigation of the pulsation constant $Q$ \citep{Breger1975}, as described in Eq.~(2).

Radial oscillations can be estimated by the pulsation constant, $Q$, as given in following,

\begin{equation}
{\rm log} Q = {\rm log} P + \frac{1}{2} {\rm log} g + \frac{1}{10} M_{\rm bol} + {\rm log} T_{\rm eff} - 6.454, \label{equation2}
\end{equation}

where $P$ is the pulsation period, $g$ is the surface gravity, $M_{\rm bol}$ is the absolute bolometric magnitude, and $T_{\rm eff}$ is the effective temperature. And the value of $M_{\rm bol}$ can be determined below \citep{Drilling2000},

\begin{equation}
M_{\rm bol} = 42.36 - 5 {\rm log} R/{R_\odot} - 10{\rm log} T_{\rm eff}, \label{equation3}
\end{equation}

where $R$ is the stellar radius and ${R_\odot}$ the solar radius, respectively. Therefore, the determination of the $Q$ value depends on the accuracy of stellar parameters such as $T_{\rm eff}$, log $g$, and $M_{\rm bol}$. In this work, for KIC 9845907, \tess mission provides the parameters as $T_{\rm eff}$ = 7861.0 ± 147 K, $R$ = 1.928 ± 0.067 $R_{\odot}$ and log $g$ = 4.140 ± 0.077 dex \citep{Stassun2019}. Using these physical parameters, we found that the pulsation constant ($Q$) for $f_{1}$ was determined to be 0.029 ± 0.002, which is in the range of values for the fundamental and first overtone modes obtained by \cite{Lovekin2017} in a grid of rotating models representative of $\delta$ Scuti stars. Hence, we suggest $f_{1}$ is a radial mode (and further identified to be the first overtone in Sec.~4.6), as marked in Table~A1 in the Appendix.

The frequencies $f_{2}$ $\sim$ $f_{7}$ are within the frequency range of typical $\delta$ Scuti stars. They are neither combination nor harmonic frequencies, so we marked them as independent frequencies in Table~A1 in the Appendix. Fig.~\ref{fig:real_frequencies} shows four frequencies ($f_{1}$, $f_{3}$, $f_{8}$ and $f_{9}$), where $f_{1}$ and $f_{3}$ are two real frequencies. We found that the shape of $f_{8}$ and $f_{9}$ are similar to that of $f_{1}$ and $f_{3}$, with no multiplet structure around the peaks. This suggests that $f_{8}$ and $f_{9}$ are not instrumental artifacts. For more detailed description about the real and alias frequencies of the \kepler data, we refer to \cite{Murphy2013}. Moreover, the S/N of $f_{8}$ and $f_{9}$ in SC spectrum are 28.7 and 20.0 respectively, which is far greater than the standard of S/N = 4 given by \cite{Breger1993}. Hence, $f_{8}$ and $f_{9}$ are two real frequencies.

\begin{figure*}[h]
\begin{center}
  \includegraphics[width=1.00\textwidth,trim=45 10 55 10,clip]{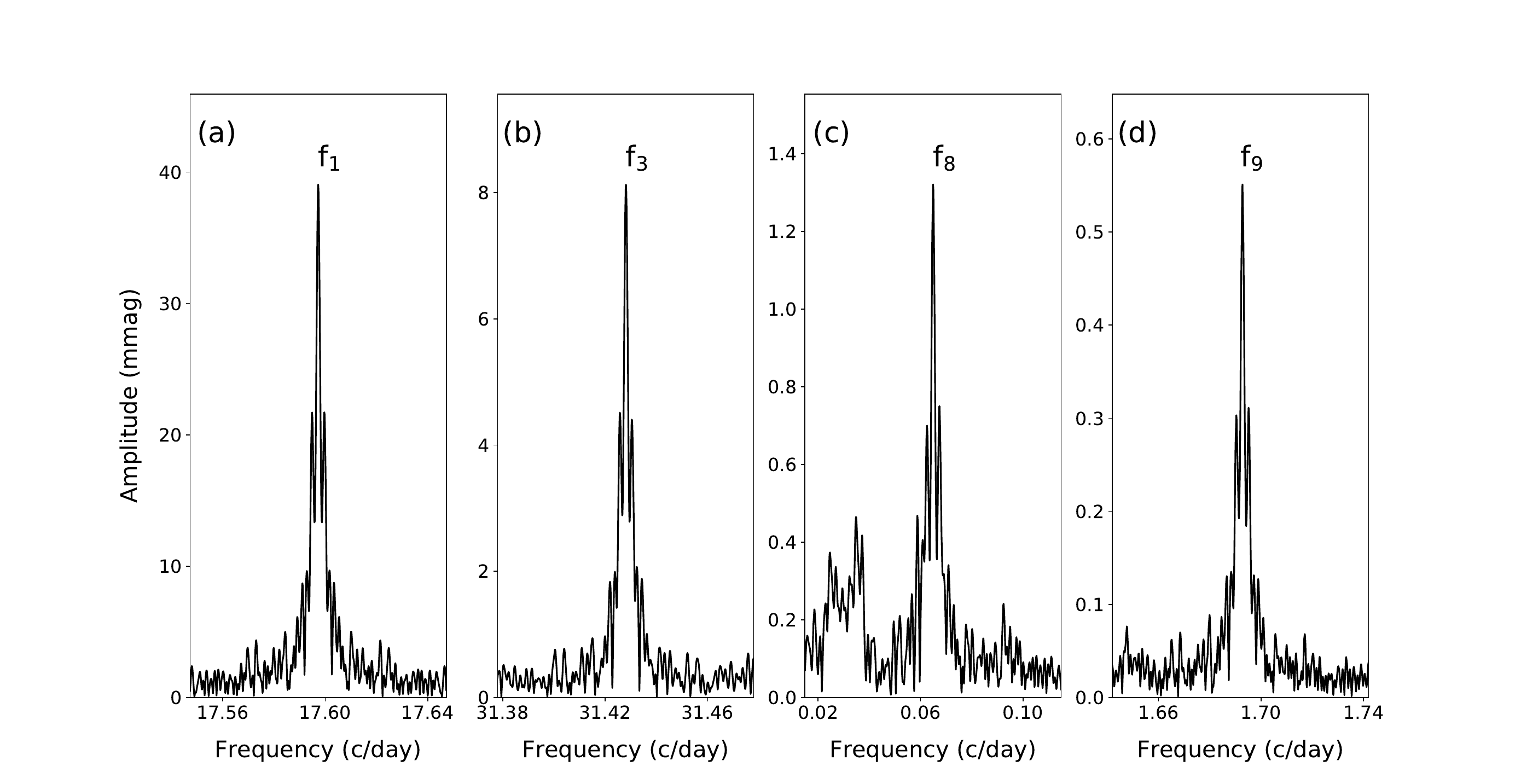}
  \caption{Typical spectra of the real frequencies. Panels (a) and (b) show two real frequencies $f_{1}$ = 17.597 day$^{-1}$ and $f_{3}$ = 31.428 day$^{-1}$; panels (c) and (d) show two modulation terms $f_{8}$ = 0.065 day$^{-1}$ and $f_{9}$ = 1.693 day$^{-1}$, respectively.}
    \label{fig:real_frequencies}
\end{center}
\end{figure*}

\begin{figure*}[h]
\begin{center}
  \includegraphics[width=1.00\textwidth,trim=45 0 55 30,clip]{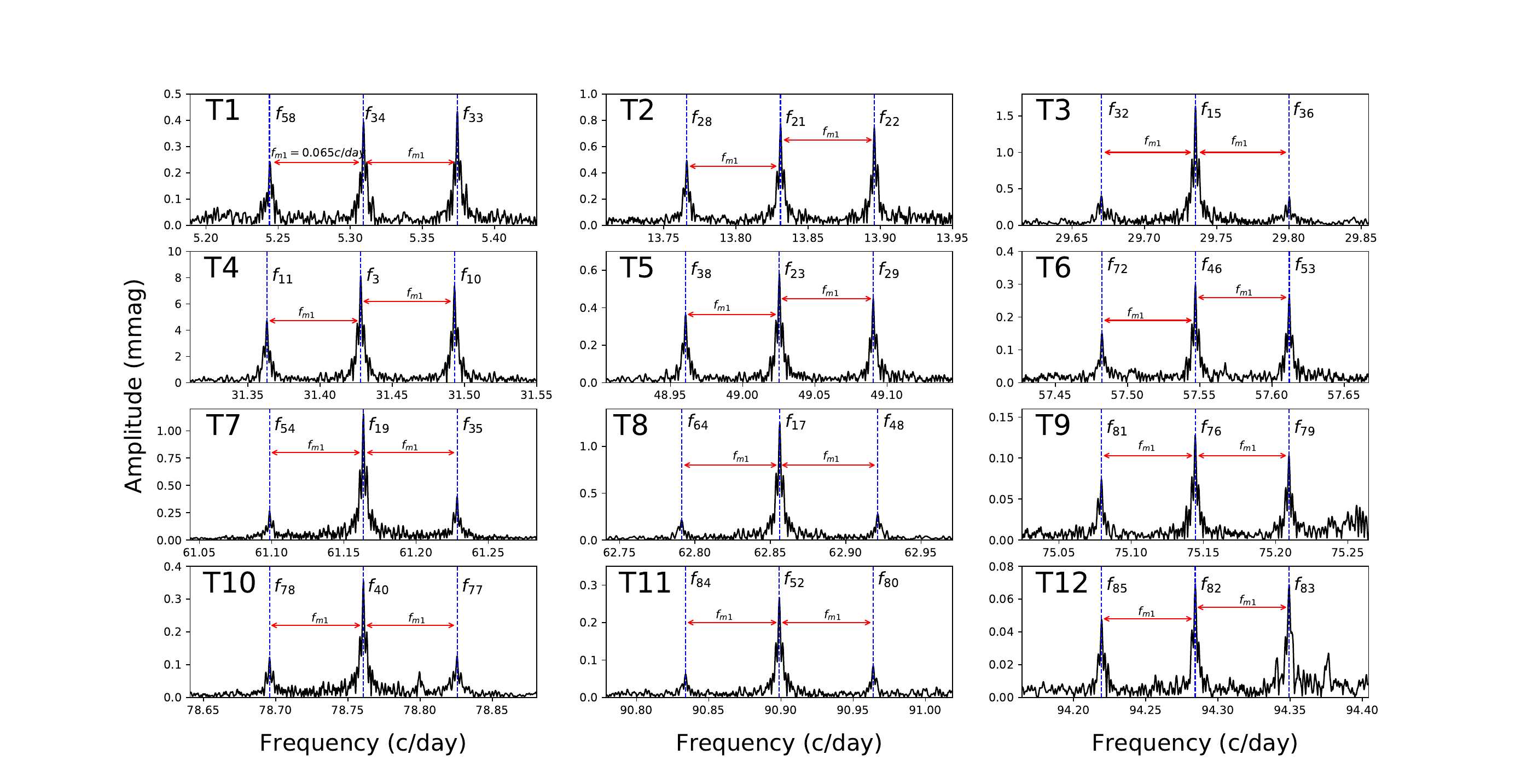}
  \caption{Twelve pairs of equidistant frequency-triplet structures in the SC spectrum of KIC 9845907. The vertical blue dotted lines indicate the locations of the detected frequencies, where $f_{m1}$ = 0.065 day$^{-1}$ represents the spacing. We marked the triplets with $f_{m1}$ as T1 to T12 sorted according to the increasing value of the central frequency.}
    \label{Triple1}
\end{center}
\end{figure*}

\begin{figure*}[h]
\begin{center}
  \includegraphics[width=1.00\textwidth,trim=45 0 55 30,clip]{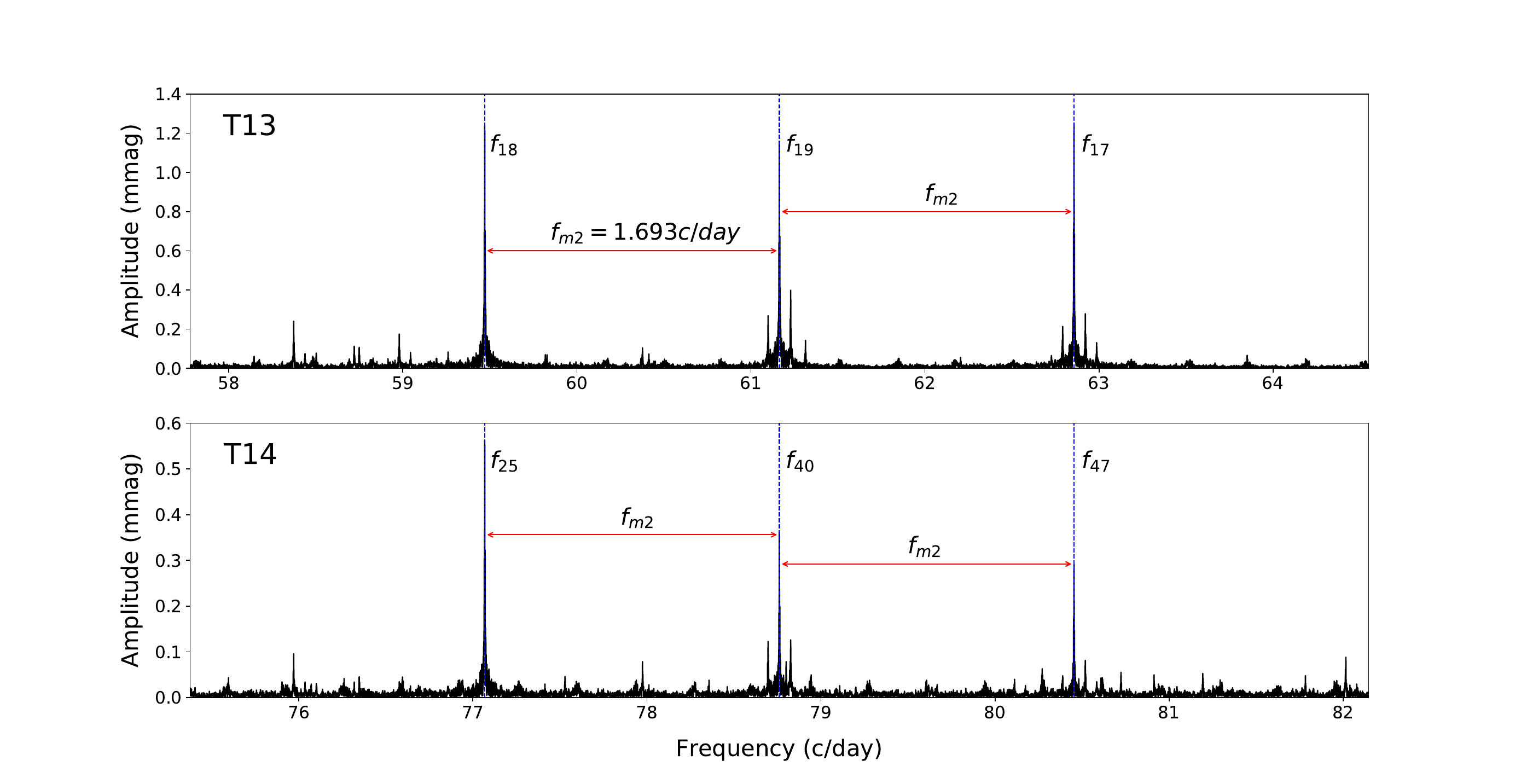}
  \caption{Two pairs of equidistant frequency-triplet structures in SC spectrum of KIC 9845907. The vertical blue dotted lines indicate the locations of frequencies, where $f_{m2}$ = 1.693 day$^{-1}$ represents the equally spaced frequency. We marked the triplets with $f_{m2}$ as T13 and T14 according to Table \ref{tab:T}.}
    \label{Triple2}
\end{center}
\end{figure*}

It is intriguing that 42 of the 85 frequencies form fourteen pairs of equidistant triplet structures. They are labeled as "T" in Figs.~\ref{Triple1} and \ref{Triple2}, where the blue vertical dotted lines indicate the locations of the extracted peaks. It is clear that these equidistant frequency triplets consist of the equal spacing frequency $f_{8}$ (= 0.065 day$^{-1}$) in Fig.~\ref{Triple1} and $f_{9}$ (= 1.693 day$^{-1}$) in Fig.~\ref{Triple2}, respectively. Moreover, $f_{8}$ and $f_{9}$ are not in the typical $p-$mode range of $\delta$ Scuti stars, and they are not combinations of other independent frequencies, either. Hence, we marked the frequencies $f_{8}$ and $f_{9}$ as $f_{m1}$ and $f_{m2}$ in Table~A1 in the Appendix, respectively. In Table \ref{tab:T}, we marked the triplets with the equidistant frequency $f_{m1}$ as T1 to T12 and the group with $f_{m2}$ as T13 and T14,  sorted according to the increasing value of the central frequency, respectively.

\begin{deluxetable}{cccccccccc}
\tabletypesize{\small} 
\tablewidth{0pc} 
\tablecaption{Dominant Frequencies of the "T" Family in the SC Data (Denoted by $f_{i}$)
\label{tab:T}} 
\tablehead{ 
\colhead{$f_{i}$}   & 
\colhead{Frequency}  &
\colhead{Identification} & 
\colhead{Comment}   & 
\colhead{$f_{i}$}   & 
\colhead{Frequency}  &
\colhead{Identification} & 
\colhead{Comment}   \\
\colhead{}   & 
\colhead{(day$^{-1}$)}  &
\colhead{} & 
\colhead{}   & 
\colhead{}   & 
\colhead{(day$^{-1}$)}  &
\colhead{} & 
\colhead{}   & 
}
\startdata 
    8      &  0.0649(1)  & $f_{8}$             &  $f_{m1}$  &          9      &  1.6927(5)  & $f_{9}$      &  $f_{m2}$ & \\     
    34     &  5.3091(2)  & $f_{34}$            &  T1        &          17     &  62.8562(6) & $f_{17}$     &  T8        &  \\
    33     &  5.3741(4)  & $f_{34}$+$f_{m1}$   &  T1        &          48     &  62.9211(7) & $f_{17}$+$f_{m1}$  &  T8     &\\
    58     &  5.2443(3)  & $f_{34}$$-$$f_{m1}$ &  T1        &          64     &  62.7913(6) & $f_{17}$$-$$f_{m1}$  &  T8    & \\
    21     &  13.8308(6) & $f_{21}$            &  T2        &          76     &  75.1444(1) & $f_{76}$      &  T9        & \\
    22     &  13.8957(6) & $f_{21}$+$f_{m1}$   &  T2        &          79     &  75.2092(7) & $f_{76}$+$f_{m1}$  &  T9   &      \\
    28     &  13.7659(5) & $f_{21}$$-$$f_{m1}$ &  T2        &          81     &  75.0793(9) & $f_{76}$$-$$f_{m1}$  &  T9  &       \\
    15     &  29.7353(7) & $f_{15}$            &  T3        &          40     &  78.7607(8) & $f_{40}$      &  T10         &\\   
    36     &  29.8002(8) & $f_{15}$+$f_{m1}$   &  T3        &          77     &  78.8256(8) & $f_{40}$+$f_{m1}$  &  T10   &      \\  
    32     &  29.6703(6) & $f_{15}$$-$$f_{m1}$ &  T3        &          78     &  78.6958(8) & $f_{40}$$-$$f_{m1}$  &  T10   &      \\  
    3      &  31.42813(3)& $f_{3}$             &  T4        &          52     &  90.8987(7) & $f_{52}$      &  T11        & \\
    10     &  31.4930(3) & $f_{3}$+$f_{m1}$    &  T4        &          80     &  90.9639(1) & $f_{52}$+$f_{m1}$  &  T11    &     \\
    11     &  31.3631(1) & $f_{3}$$-$$f_{m1}$    &  T4        &          84     &  90.8339(8) & $f_{52}$$-$$f_{m1}$  &  T11   &      \\
    23     &  49.0254(5) & $f_{23}$            &  T5        &          82     &  94.2843(9) & $f_{82}$     &  T12       & \\
    29     &  49.0903(1) & $f_{23}$+$f_{m1}$   &  T5        &          83     &  94.3493(4) & $f_{82}$+$f_{m1}$ &  T12     &   \\
    38     &  48.9605(1) & $f_{23}$$-$$f_{m1}$   &  T5        &          85     &  94.2194(9) & $f_{82}$$-$$f_{m1}$ &  T12      &  \\
    46     &  57.5470(3) & $f_{46}$            &  T6        &          19     &  61.1635(1) &  $f_{19}$       &  T13       & \\
    53     &  57.6120(4) & $f_{46}$+$f_{m1}$   &  T6        &          17     &  62.8562(6) & $f_{19}$+$f_{m2}$  &  T13  &      \\
    72     &  57.4821(2) & $f_{46}$$-$$f_{m1}$   &  T6        &          18     &  59.4707(5) & $f_{19}$$-$$f_{m2}$  &  T13  &      \\
    19     &  61.1635(1) & $f_{19}$            &  T7        &          40     &  78.7607(8) & $f_{40}$      &  T14        &\\
    35     &  61.2284(1) & $f_{19}$+$f_{m1}$   &  T7        &          47     &  80.4535(8) & $f_{40}$+$f_{m2}$  &  T14   &     \\
    54     &  61.0984(9) & $f_{19}$$-$$f_{m1}$   &  T7        &          25     &  77.0680(3) & $f_{40}$$-$$f_{m2}$  &  T14    &    \\
    
   \enddata 
    \tablecomments{"T", $f_{m1}$ and $f_{m2}$ represent equidistant frequency-triplet structures and two equally spaced frequencies, respectively. We marked the equidistant frequency triplets of the group with $f_{m1}$ as T1 to T12 and the group with $f_{m2}$ as T13 and T14, sorted according to the increasing value of the central frequency, respectively.}
\end{deluxetable}  

\begin{figure*}[h]
\begin{center}
  \includegraphics[width=1.00\textwidth,trim=45 0 55 30,clip]{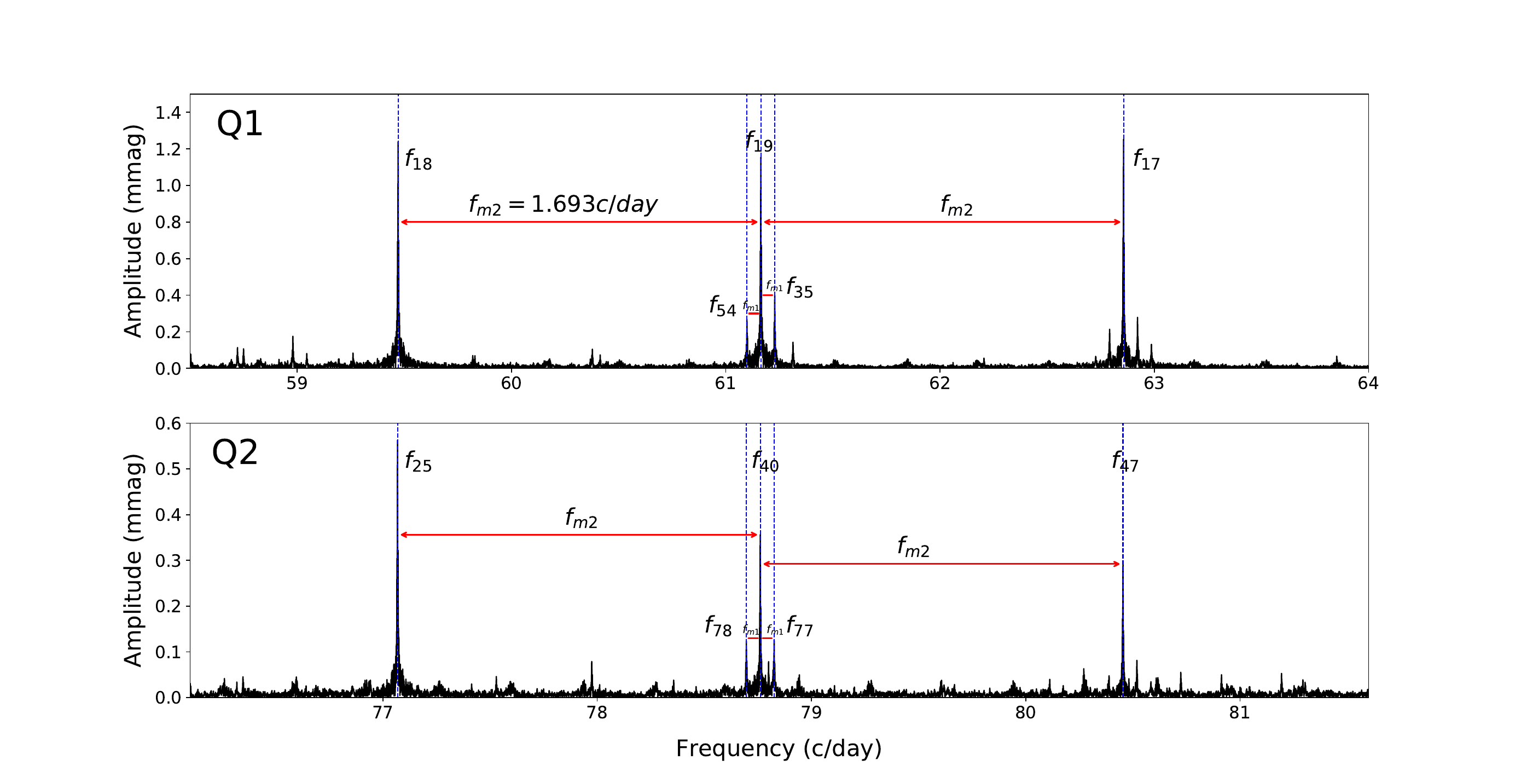}
  \caption{Two pairs of quintuplet structures in SC spectrum of KIC 9845907. The vertical blue dotted lines indicate the locations of frequencies. We marked the two pairs as Q1 (upper panel) and Q2 (lower panel), sorted according to the increasing value of the central frequency, respectively. The intervals from the side peaks to the center are marked with $f_{m1}$ = 0.065 day$^{-1}$ and $f_{m2}$ = 1.693 day$^{-1}$, respectively. In the upper and lower panels, the zoomed-in view of the central triplets can be seen in Fig.~\ref{Triple1} (i.e., T7 and T10) for clarity.}
    \label{fig:Q}
\end{center}
\end{figure*}

Moreover, we also found two pairs of quintuplet structures ("Q") in the SC frequency spectra. In Table \ref{tab:Q}, we marked the two pairs as Q1 and Q2, sorted according to the increasing value of the central frequency. Fig.~\ref{fig:Q} shows these two pairs of quintuplet structure with the blue dotted lines representing the location of each extracted frequency. It is clear that the side peaks around $f_{19}$ (and $f_{40}$) in the frequency spectra of KIC 9845907 have two pairs of uniformly spaced triplets with intervals of $f_{m1}$ = 0.065 day$^{-1}$ and $f_{m2}$ = 1.693 day$^{-1}$, i.e., Q1 includes T7 and T13, while Q2 includes T10 and T14, respectively. To see clearly the central triplets with $f_{m1}$ in Fig.~\ref{fig:Q}, we refer to T7 and T10 in Fig.~\ref{Triple1}. Note that four out of fourteen pairs of triplets can form quintuplet structures, which is very interesting and recognized for the first time in $\delta$ Scuti stars. We suggest these phenomena deserve further study, and we give a discussion on the possible explanations in Section 4.

\begin{deluxetable}{ccccccccccc} 
\tabletypesize{\small} 
\tablewidth{0pc} 
\tablecaption{Dominant Frequencies of the "Q" Family in the SC Data (Denoted by $f_{i}$)}
\label{tab:Q}
\tablehead{ 
\colhead{$f_{i}$}   & 
\colhead{Frequency}  &
\colhead{Identification} & 
\colhead{Comment}   & 
\colhead{$f_{i}$}   & 
\colhead{Frequency}  &
\colhead{Identification} & 
\colhead{Comment}   \\
\colhead{}   & 
\colhead{(day$^{-1}$)}  &
\colhead{} & 
\colhead{}   & 
\colhead{}   & 
\colhead{(day$^{-1}$)}  &
\colhead{} & 
\colhead{}   &
}
\startdata
   19      &  61.1635(1) &  $f_{19}$           &  Q1   &  40      &  78.7607(8) &  $f_{40}$           &  Q2   & \\
   17      &  62.8562(6) &  $f_{19}$+$f_{m2}$  &  Q1   &  47      &  80.4535(8) &  $f_{40}$+$f_{m2}$  &  Q2   &  \\
   35      &  61.2284(1) &  $f_{19}$+$f_{m1}$  &  Q1   &  77      &  78.8256(8) &  $f_{40}$+$f_{m1}$  &  Q2   &  \\
   54      &  61.0984(9) &  $f_{19}$$-$$f_{m1}$  &  Q1   &  78      &  78.6958(8) &  $f_{40}$$-$$f_{m1}$  &  Q2   &   \\
   18      &  59.4707(5) &  $f_{19}$$-$$f_{m2}$  &  Q1   &  25      &  77.0680(3) &  $f_{40}$$-$$f_{m2}$  &  Q2   &    \\       
         
   \enddata 
    \tablecomments{"Q" represents quintuplet structures. We marked the two pairs as Q1 and Q2, sorted according to the increasing value of the central frequency, respectively.}
\end{deluxetable}

\section{Discussion}

In recent years, more and more regular frequency spacings have been found in several $\delta$ Scuti stars, thanks to the high-precision continuous photometric data provided by space telescopes. The identification of these pulsation frequency modes provides us with new clues to study these variable stars. For example, \cite{Bedding2020} studied 60 $\delta$ Scuti stars with regular frequency spacings and enabled mode identification by comparing with theoretical models. 

The most interesting features for KIC 9845907 in this study are the regular frequency spacings, including equidistant frequency-triplet and quintuplet structures, with frequency intervals of $f_{m1}$ = 0.065 day$^{-1}$ and $f_{m2}$ = 1.693 day$^{-1}$. Two pairs of triplet structure overlap to form a quintuplet structure, hence, with the same frequency spacings (of $f_{m1}$ and $f_{m2}$). To explore the nature of these frequency structures, we first need to consider whether they arise from the known instrumental effects of \kepler, including: (1) the frequency of \kepler orbital period $f_{\rm orb}$ = 0.00268 day$^{-1}$, (2) the momentum desaturation of the reaction wheel $f_{\rm reaction}$ = 0.336 day$^{-1}$ \citep{2016ksci.rept....1V}, (3) the frequency induced by the data downlink $f_{\rm downlink}$ = 0.031 day$^{-1}$, (4) that of \kepler rotation $f_{\rm rot}$ = 0.011 day$^{-1}$. The Rayleigh frequency resolution of the SC data is 0.001136 day$^{-1}$ for KIC 9845907. We found that the value of $f_{m1}$ was not equal to (1) and (2) within the Rayleigh frequency resolution; therefore, these two instrumental effects are excluded firstly. For the frequency $f_{\rm downlink}$ (=0.031 day$^{-1}$), $f_{m1}$ is about twice of it. To verify if $f_{m1}$ is caused by the data downlink, we only used a part of 32 day (=1/$f_{\rm downlink}$) data to extract frequencies and still detected the $f_{m1}$, so this instrumental effect is excluded. For the frequency $f_{\rm rot}$, $f_{m1}$ seems to be 6 times of it. If the equidistant structure in the spectra is caused by the instrument effect, the height of the side lobe frequencies on both sides of the peak should be the same. But the triplets with $f_{m1}$ (T1$-$T12) of KIC 9845907 are not like this. In addition, instrumental effect should be applied at all significant independent frequencies. For KIC 9845907, we only detected the equidistant frequency-triplet structure at one independent mode (i.e., $f_{3}$) and none at the others. And when using the software PERIOD04 for frequency extraction, we did not detect 1$-$5 times of the \kepler rotation frequency; these signals should be stronger than 6$f_{\rm rot}$. Hence, $f_{m1}$ $\approx$ 6$f_{\rm rot}$ is probably a coincidence. For $f_{m2}$, we found that none of the known frequencies of instrumental effects from \kepler were equal to it within the Rayleigh frequency resolution.

\subsection{Beating?}

\cite{Breger2002} found that pairs of close-frequency pulsation modes with spacings less than 0.01 day$^{-1}$ were common by studying seven well-known $\delta$ Scuti stars. \cite{Breger2009} found that pulsation frequencies in $\delta$ Scuti stars are not distributed randomly and that many non-radial modes had frequencies near radial mode frequencies. These regularities were explained by mode trapping in stellar envelope \citep{Dziembowski1990}, which explained the regularities in the amplitude spectrum in $\delta$ Scuti star FG Vir \citep{Breger2009}.

However for KIC 9845907, the spacings of the close frequencies are $f_{m1}$ = 0.065 day$^{-1}$ and $f_{m2}$ = 1.693 day$^{-1}$, both of which are much larger than the typical value (i.e., 0.01 day$^{-1}$) in \cite{Breger2002}. Moreover, there is no unresolved peak around the dominant frequencies in the residuals, and the overall distribution of the residuals is typical of noise. Hence, the equidistant triplet structures of KIC 9845907 are unlikely caused by beating.

\subsection{Blazhko Effect?}

Equidistant structures are often shown in the Fourier spectra of the Blazhko RR Lyrae stars, and the interval of the triplets is the same as the modulation frequency, which can be directly detected \citep{Jurcsik2005,Kolenberg2006}. For KIC 9845907 in our case the equidistant quintuplet structures in the SC spectrum are similar to that in Blazhko RR Lyr stars, and two modulation frequencies are detected clearly. These features imply that the quintuplet structures in KIC 9845907 may be related to the Blazhko effect.

Using the high-precision photometric data provided by \kepler, Blazhko-like effect has been found in some $\delta$ Scuti stars. For example, the double-mode HADS star KIC 10284901 shows two pairs of quintuplet structure, and its analysis suggests that the modulation term of the quintuplet structures might be related to the Blazhko effect \citep{Yang2019}. The main characteristic of this effect is that the quintuplet structures appear around the fundamental and first overtone pulsation modes (i.e., F0 and F1). However, for KIC 9845907, the two pairs of quintuplet structures are not the case, that is to say, they appear around other modes, which are not similar to that in KIC 10284901. Moreover, in KIC 10284901 the ratio of the two modulation frequencies ($f_{m1}$ and $f_{m2}$) is nearly 1:2. However, in the case of KIC 9845907, the ratio of the two modulation frequencies ($f_{m1}$ = 0.065 day$^{-1}$ and $f_{m2}$ = 1.693 day$^{-1}$) of the quintuplet structures is seriously deviated from 1:2. From these aspects, we rule out the possibility that the quintuplet structures are caused by the Blazhko effect, as observed in RR Lyrae stars.

\subsection{Combination Mode Hypothesis?}

For the equidistant frequency-triplet structures in $\delta$ Scuti stars, \cite{Breger2006} provided an explanation named the "Combination Mode Hypothesis." In this hypothesis, the highest amplitude mode $v_{1}$ and a real second mode $v_{2}$ are excited; then, the harmonics as well as the combinations of $v_{1}$ and $v_{2}$ may also occur. The simplest combination (e.g., $v_{3}$ = $v_{2}$ $-$ $v_{1}$) is usually observed, and then a pair of frequency-triplet (i.e., $v_{3}$, $v_{1}$, and $v_{2}$) is formed around $v_{1}$ in the frequency spectra.

To test this possibility, we chose the triplet structure T4 (i.e., $f_{3}$, $f_{10}$ and $f_{11}$) in KIC 9845907 for further analysis since the amplitude of the central component (i.e., $f_3$) of T4 is the strongest. Under this hypothesis, the right component $f_{10}$ = 31.493 day$^{-1}$ in the equidistant frequency-triplet structure in the SC spectrum of KIC 9845907 is considered as a new independent mode. However, the ratio of $f_1$/$f_{10}$ = 0.559 is not within the typical range of the continuous radial period ratios \citep{Stellingwerf1979}. This seems to rule out the possibility that $f_{10}$ belongs to a radial mode. If $f_{10}$ is assumed to be a non-radial mode, it would split into 2$\ell$+1 frequencies when the star rotates. However, it is not like this case, either, as $f_{10}$ does not exhibit any splitting structure in the frequency spectra. Consequently, $f_{10}$ in KIC 9845907 is not an independent mode.

\subsection{Nonlinear Mode Coupling?}

Different nonlinear effects can generate combination frequencies in the spectrum of pulsating stars \citep{Bowman2016}. Combination frequencies are common in $\delta$ Scuti stars, and it is of great importance to identify which peak is combination frequency and which is a real frequency as it can greatly simplify the frequency spectra \citep{Kurtz2015}. However, combination frequencies differ from the mode coupling, which are excited by the resonant interaction of pulsation modes in stars. Coupled frequencies are grouped into two families: child and parent modes, and this coupling of modes promotes energy exchange between different modes of the family \citep{Nowakowski2005}. The frequencies and amplitudes of pulsating modes change with time due to nonlinear mode coupling. Hence, when we distinguish coupled modes from combination frequencies, we need to study the changes of frequency, amplitude, and phase for each family.

In the theory of nonlinear mode coupling, the child and parent modes are directly related in frequency and phase. The way to distinguish the parent and coupled modes is to examine the correlations in amplitude variability. \cite{Breger2014} investigated the dominant modes excited in KIC 8054146, which they called the T family. They computed changes of amplitudes and phases and correlations between different frequencies that show qualitative similarities between Triplets 1 and 2 (see the Figs.~2 and 3 in \citealt{Breger2014}). Finally, they identified the coupled modes of the T family in KIC 8054146. In this work, following the method by \cite{Breger2014}, we investigated the correlations in amplitude and phase variability between different dominant pairs of equidistant triplet structures.

\begin{figure*}[h]
\begin{center}
  \includegraphics[width=1.00\textwidth,trim=45 30 55 30,clip]{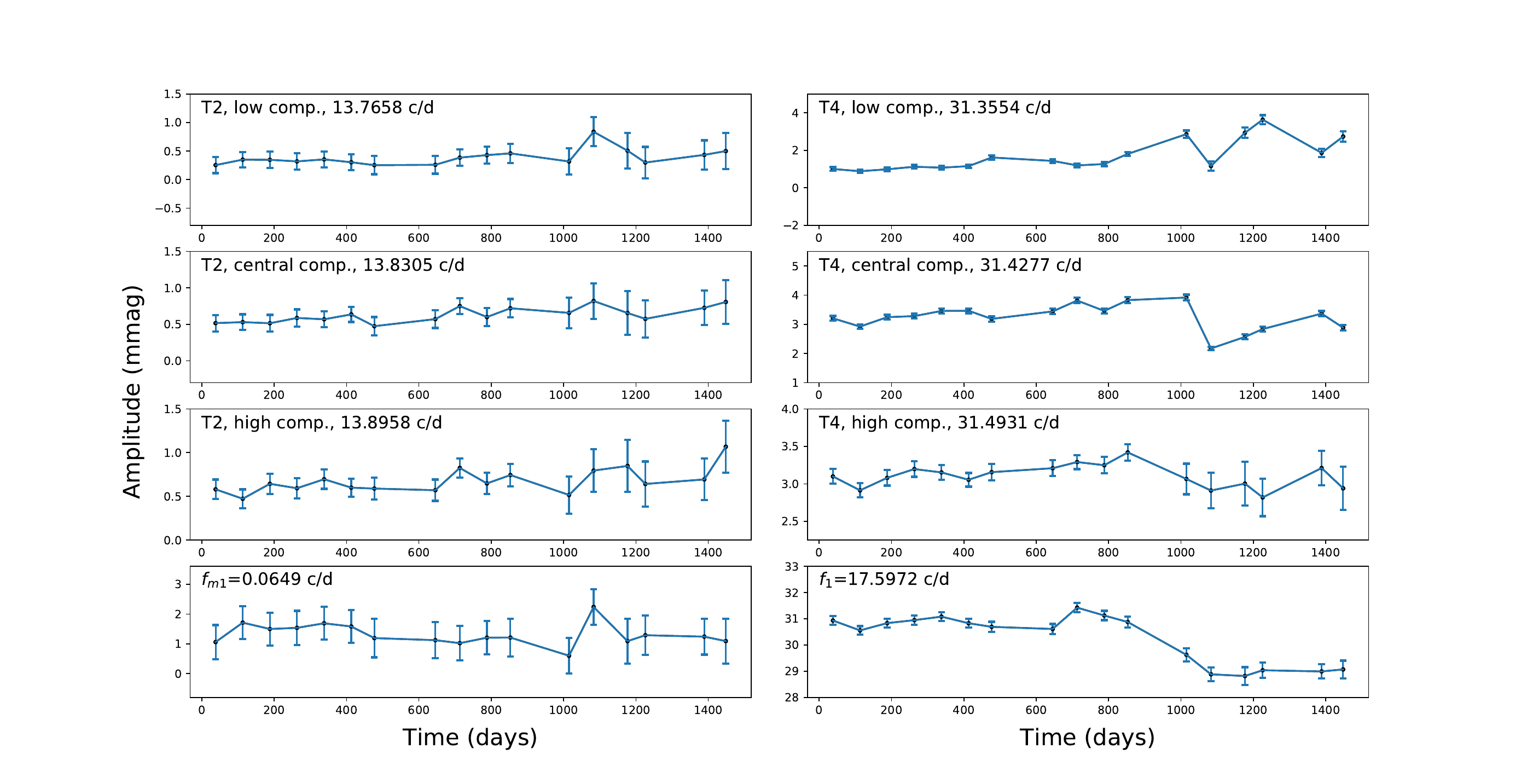}
  \caption{Amplitude variations of the components of Triplet 4 (T4) and Triplet 2 (T2). Two dominant frequencies, $f_{m1}$ and $f_{1}$, are also shown for comparison.}
    \label{fig:amp_variations}
\end{center}
\end{figure*}

\begin{figure*}[h]
\begin{center}
  \includegraphics[width=1.00\textwidth,trim=45 30 55 50,clip]{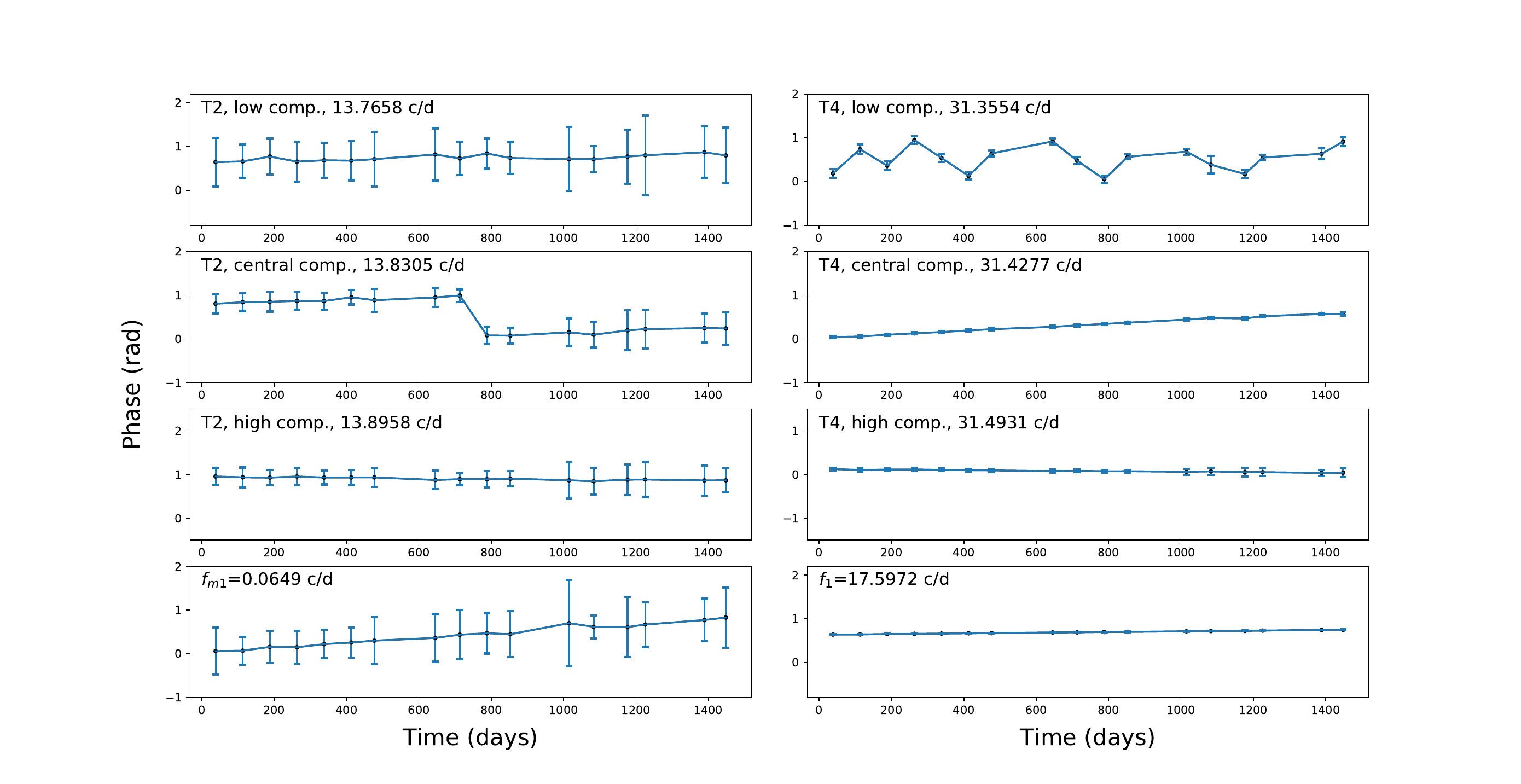}
  \caption{The same as in Fig.~\ref{fig:amp_variations} but for phase variations of the modes.}
    \label{fig:Phase_variations}
\end{center}
\end{figure*}

The 4 yr of available LC data of KIC 9845907 allow us to examine the correlations of the amplitude and phase changes of the dominant frequencies and help to detect their physical relationship and origin. This near-continuous LC data from Q0 to Q17 span 1471 days in total. To make a compromise between studying short-term amplitude and phase changes and obtaining excellent frequencies from larger time intervals, a 75 day interval was chosen. In the next step, average frequencies covering the entire Q0 $-$ Q17 quarters were determined. Each value of the frequencies is labeled in the upper left corner of each panel in Fig.~\ref{fig:amp_variations}. Using these average frequencies, the amplitudes and phase were calculated for different modes in each 75 day interval. This is shown in Fig.~\ref{fig:amp_variations} and Fig.~\ref{fig:Phase_variations}, respectively. In the last row we also show the results for modes at $f_{m1}$ = 0.065 day$^{-1}$ and $f_{1}$ = 17.597 day$^{-1}$ for comparison. The error bars are also plotted, obtained by Monte Carlo simulation in PERIOD04.

Fig.~\ref{fig:amp_variations} shows the amplitude variations of KIC 9845907. The panels in the top three rows represent the low, central, and high components (from top to bottom) of T2 (left) and T4 (right), respectively. The bottom panels are the amplitude variations of $f_{m1}$ and $f_{1}$, respectively. For the three components of T2, before 1000 days, the amplitude variations were relatively stable but then changed significantly (i.e., increasing or decreasing). However, for T4, the changes occur much earlier, at the time of 700 days. For $f_{m1}$, the amplitude change is relatively stable as a whole without obvious changes, and the frequency $f_{1}$ breaks the stability after 700 days. Then it tends to be stable after a significant decline. We note there is no similarity and synchronization of amplitude variations between T2 and T4, which is not similar to the case of KIC 8054146.

Fig.~\ref{fig:Phase_variations} shows the phase variations of KIC 9845907 with the panels labeled the same as in Fig.~\ref{fig:amp_variations}. For the low and high components in T2, there is no obvious change in the overall phase, but the central component has a significant decline after 600 days. For the phase variations of T4, the three components have different changes: the low component can increase and/or decrease, while the central is gradually increasing, and the high is relatively stable. $f_{m1}$ and $f_{1}$ are in a slowly increasing and stable state, respectively. Therefore, there is no similarity and synchronization between T2 and T4, either.

We also made a further quantitative calculation following the method provided by \citet[][see their Eq.~(5)]{Breger2014} and confirmed that T4 is not the result of mode coupling between T2 and the radial mode $f_{1}$. We conclude that there is no sufficient evidence so far to support the possibility that the triplets are caused by nonlinear mode coupling.

\subsection{Large Separation?}

Observations have shown that many $\delta$ Scuti stars have regular frequency spacings in their rich pulsation spectra. There have been several studies searching for large separations (e.g., \citealt{2013A&A...559A..63G,Bedding2020}). This development follows the establishment of the relationship between the large separation in the low-order regions and the average density of the star from the modelings (e.g., \citealt{2014A&A...563A...7S}) and the observations \citep{2015ApJ...811L..29G,2017MNRAS.471L.140G}.

\begin{figure*}[h]
\begin{center}
  \includegraphics[width=1.00\textwidth,trim=20 30 30
   30,clip]{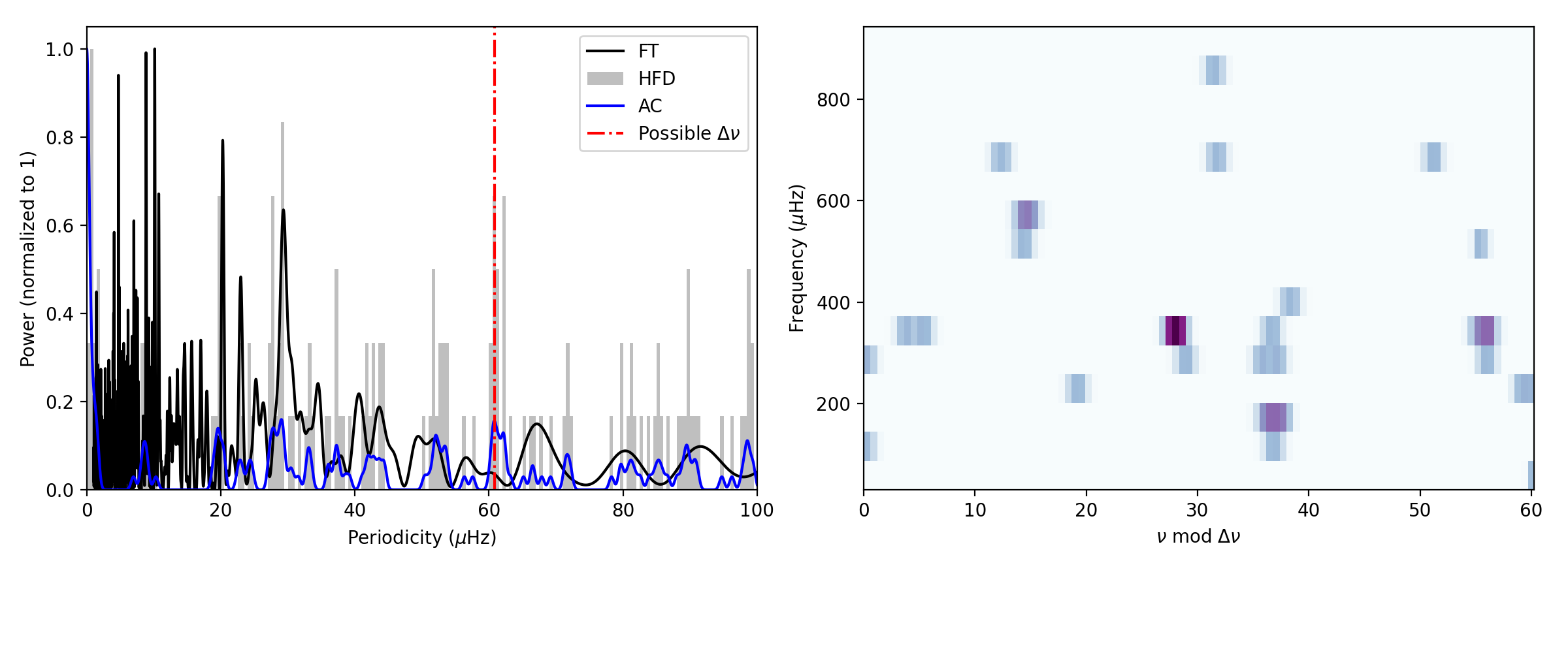}
  \caption{In the left panel, the black, gray, and blue lines represent the Fourier transform (FT), the histogram of frequency differences (HFD), and the autocorrelation function (AC), respectively, using the methodology of \cite{2013A&A...559A..63G} and \cite{Ram2021}. The red dotted$-$dashed line indicates the possible value of $\Delta \nu$ = 5.25 day$^{-1}$. The right panel shows the echelle diagram of the frequencies. The value of 60.8 $\mu$Hz is used for the plot.}
    \label{fig:Dnu}
\end{center}
\end{figure*}

We used the methodology provided by \cite{2013A&A...559A..63G} and \cite{Ram2021} to determine the large separation, $\Delta \nu$, of KIC 9845907. Three techniques were applied to the frequencies: (1) the FT, (2) the autocorrelation function (AC), (3) the histogram of frequency differences (HFD). The result of KIC 9845907 is shown in the left panel of Fig.~\ref{fig:Dnu}, where the black, gray, and blue lines represent FT, HFD, and AC, respectively. All these transformations have been made using just the 30 highest amplitude modes and discarding those frequencies below 5 day$^{-1}$, and the amplitudes of the frequencies have been normalized to unity when doing all these transformations. If there exists a periodicity in the form of Dirac comb, then we will see a peak corresponding to $\Delta \nu$ and others at the multiples in the AC (2$\Delta \nu$, 3$\Delta \nu$, etc). On the contrary, in the FT we expect to find $\Delta \nu$ and the submultiples ($\Delta \nu$/2, $\Delta \nu$/3, etc), just because of how the transform works. In conclusion, in the FT, we could not find $\Delta \nu$ but just the submultiples ($\Delta \nu$/2 and/or $\Delta \nu$/3) while $\Delta \nu$ appears in the AC and HFD.

For KIC 9845907, the left panel of Fig.~\ref{fig:Dnu} shows a peak at around 5.25 day$^{-1}$ (= 60.8 $\mu$Hz, the red dotted$-$dashed line) in the AC and HFD. However, in the FT, it does not show a peak around $\Delta \nu$, but we can see $\Delta \nu$/2 (the high peak around 30 $\mu$Hz) and $\Delta \nu$/3 (the high peak around 20 $\mu$Hz). The absence of the peak of $\Delta \nu$ in the FT is expected, although to identify it at least one submultiple is mandatory (e.g., \citealt{2009A&A...506...79G}), and the presence of more submultiples indicates that the possible value of large separation ($\Delta \nu$) is around 60.8 $\mu$Hz and not 30 $\mu$Hz. Moreover, the large separation ($\sim$ 5.1 day$^{-1}$) we obtained from the models (see Sec. 4.6 for more detail) is close to the observed one.

Using this value (60.8 $\mu$Hz), the echelle diagram is shown in the right panel of Fig.~\ref{fig:Dnu}. Some frequencies appear darker as they are composed of two very close frequencies. It can be seen that the distribution of these modes is irregular, with additional ridges at various angles, indicating that the analysis method of using simple pulsation spectra (with $\ell$ = 0 and 1) is insufficient to interpret this echelle diagram. In fact, this phenomenon also appears in other $\delta$ Scuti stars. For example, in four $\delta$ Scuti stars (i.e., HD 37286, SAO 150524, TYC 8533-329-1, and $\beta$ Pic), complex patterns were also detected in the echelle diagrams. \cite{Bedding2020} concluded that the complex echelle diagram might be related to the azimuthal order $m$. Therefore, only using the large separation 60.8 $\mu$Hz cannot provide a reasonable explanation for the triple structures and the rich spectrum for KIC 9845907.

\subsection{Stellar Rotational Splitting?}

The theory of stellar oscillations in the spherical approximation states that each oscillation mode can be characterized by three spherical harmonics: number of nodes along the radius direction $n$, the spherical harmonic degree $\ell$ and the azimuthal order $m$. If a star is rotating, the rotation can cause the non-radial oscillation mode to split into 2$\ell$+1 components in the inertial frame. And at first order in the perturbation theory, in a slowly rotating star, the 2$\ell$+1 frequencies of the non-radial mode are separated by almost the same spacing, corresponding to an average rotation frequency. In addition, if a star is in a binary system, the binarity would split all the modes, even the radial, because of the motion around the center of masses \citep{Shibahashi2012}. This is not the case for KIC 9845907, as the frequency $f_{1}$, a radial order and the highest amplitude peak, did not split, implying that this star should be a single star. Hence, the rotational splitting of a single star was considered to account for the triplet structures in KIC 9845907. In T1$-$T14, the highest amplitude of the middle peak is $f_{3}$ (in T4), which is an independent frequency, while all other middle peaks are combination frequencies, so only T4 will be discussed in detail in this section. To verify whether T4 is caused by the rotational splitting of $f_{3}$, we used asteroseismic models to estimate the value of $l$ for $f_{3}$.

We constructed a grid of evolutionary models of KIC 9845907 and calculated their corresponding frequencies using the submodule $``pulse{\_}adipls"$ of the Modules for Experiments in Stellar Astrophysics (MESA v10398; \citealt{Paxton2011,Paxton2013,Paxton2015,Paxton2018,Paxton2019}). In the grid, the stellar mass ranges from 1.5 $M_{\odot}$ to 2.0 $M_{\odot}$ with a step of 0.01 $M_{\odot}$, and metallicities range from 0.006 to 0.015 with a step of 0.001. For the helium abundance $Y$, we adopted $Y$ = 0.249 + 1.33 $Z$ as a function of $Z$. In addition, the classical mixing length theory of \cite{1958ZA.....46..108B} with $\alpha$ = 1.9 \citep{Paxton2013} was adopted. Each model in the above grid was evolved from the zero-age MS to the post-MS stage to calculate the pulsation frequencies of $\ell$ = 0, 1, and 2. Then, by using the method from \citet[][i.e., Eq. 5, $\chi^2$ method]{Chen2019}, the goodness of fit can be obtained by comparing model frequencies with the observed frequencies $f_{1}$ and $f_{3}$. To select the best-fitting models, we chose a threshold of $\chi^2$ = 0.0056 (corresponding to about 5 times the frequency resolution 1/$\Delta$T) and limited the model results by combining the observed effective temperature (7700 K < $T_{\rm eff}$ < 8100 K, covering the values given by \kepler and \tess) and luminosity (14.45 < $L$/$L_{\odot}$ < 15.85, estimated from $Gaia$ parallax). Finally, four candidate models are obtained, as described in the following.

\begin{figure*}[h]
\begin{center}
  \includegraphics[width=1.00\textwidth,trim=45 30 55 30,clip]{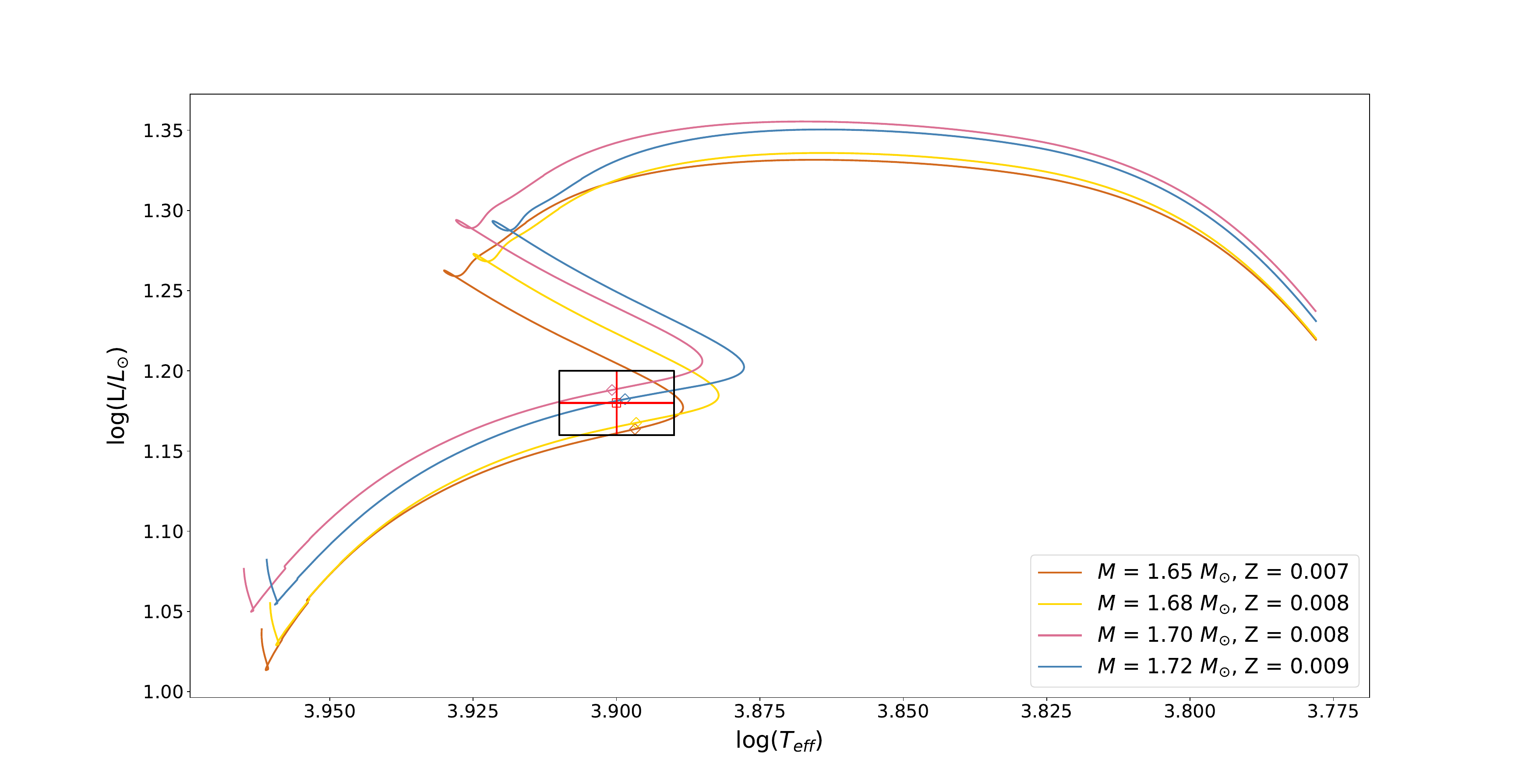}
  \caption{Evolutionary tracks from the zero-age MS to the post-MS for the four candidate models, as listed in Table 4. The black rectangle marks the 1$\sigma$ error box of the effective temperature $T_{\rm eff}$ and the luminosity $L$ while the diamonds mark the minimum $\chi^2$ for each specific model by fitting the observed $f_{1}$ = 17.597 day$^{-1}$ and $f_{3}$ = 31.428 day$^{-1}$ with the theoretical values. }
    \label{fig:hr}
\end{center}
\end{figure*}

Fig.~\ref{fig:hr} shows the evolutionary tracks of the four candidate models on the H-R diagram, where different colored lines represent different models. The black rectangle marks the 1$\sigma$ error box of the effective temperature $T_{\rm eff}$ and the luminosity $L$, while the diamonds denote the seismic models that can best reproduce the frequencies of the two modes $f_{1}$ and $f_{3}$. Table 4 lists the parameters of the four candidate models. The ($\ell, n$) of $f_{1}$ of all candidate models is (0, 2), indicating that $f_{1}$ is the first overtone, which makes KIC 9845907 a $\delta$ Scuti star with the first overtone as the dominant frequency. For $f_{3}$, the ($\ell, n$) of all candidate models is (1, 6), indicating that $f_{3}$ is a non-radial oscillation mode with $\ell$ = 1. Then, it may result into 2$\ell$+1 (i.e., 3) frequencies due to the rotational splitting, as shown in T4. Thus, we suggest equidistant triplet structures with $f_{m1}$ = 0.065 day$^{-1}$ (i.e., T1 $-$ T12) might be related to rotational splitting. For the T13 and T14 (as well as Q1 and Q2), we suggest that they are derived from combination of independent frequencies (see Table~A1 in the Appendix). The mode identification for other independent frequencies deserves a thorough investigation, which is, however, beyond the scope of this paper.

With the value of the rotational splitting (i.e., $f_{m1}$ = 0.065 day$^{-1}$), the rotational period of KIC 9845907 can be calculated, i.e., $P_{\rm rot}$ = 7.69 day. Using the formula $v_{\rm rot} = \frac{2\pi R}{P_{\rm rot}}$ and the radius of KIC 9845907 provided by \tess, $R$ = 1.928 ± 0.067 $R_{\odot}$ \citep{Stassun2019}, the rotation rate of KIC 9845907 at the equator can be obtained, i.e., $v_{\rm rot}$ ($v$${\rm sin}$$i$) = 12.70 $_{-0.46}^{+0.42}$ km $\rm s^{-1}$. Note that our result is similar to the rotational velocity (10.859 km $\rm s^{-1}$) measured by \cite{Xiang2022} using low-resolution spectra and the analysis by 
\cite{Niemczura2017} using high-resolution spectroscopy. When compared with the average rotation ($\sim$ 150 km $\rm s^{-1}$) of the typical $\delta$ Scuti stars \citep{Breger2000}, the rotation of KIC 9845907 suggests that it is a slow rotator. This is also in agreement with a symmetric structure of the split frequencies

\begin{deluxetable}{cccccccc} 
\tabletypesize{\small} 
\tablewidth{0pc} 
\tablecaption{Candidate Models with $\chi^{2} \leq 0.0056$}
\label{tab:model}
\tablehead{ 
\colhead{Mass}   & 
\colhead{$Z$}  &
\colhead{$T_{\rm eff}$} & 
\colhead{log($L$/$L_{\odot}$)}   & 
\colhead{$f_{1}$ ($\ell, n$)}   & 
\colhead{$f_{3}$ ($\ell, n$)}  &
\colhead{$\chi^{2}$}       \\
\colhead{($M_{\odot}$)}   & 
\colhead{(dex)}  &
\colhead{(K)} & 
\colhead{}   & 
\colhead{(c $\rm days^{-1}$)}   & 
\colhead{(c $\rm days^{-1}$)}  &
\colhead{} & 
}
\startdata
         1.65      &  0.007 &    7885        &  1.164   &   17.5405 (0,2)    &  31.4926 (1,6)  &      0.0037       \\
         1.68      &  0.008 &    7881        &  1.168   &   17.5386 (0,2)    &  31.4876 (1,6)  &      0.0035         \\   
         1.70      &  0.008 &    7958        &  1.188   &   17.5364 (0,2)    &  31.4752 (1,6) &      0.0031        \\
         1.72      &  0.009 &    7916        &  1.182   &   17.5170 (0,2)    &  31.4582 (1,6) &      0.0037         \\            
  \enddata 
   \tablecomments{($\ell, n$) are the spherical harmonic degree and the radial order of the model frequency, respectively.}
\end{deluxetable}

\section{SUMMARY}

Based on the high-precision time-series photometric SC data from the \kepler, we analyzed the pulsations of KIC 9845907 and detected 85 significant frequencies (see Fig.~2 and Table~A1 in the Appendix), including the radial frequency $f_{1}$ = 17.597 day$^{-1}$, the non-radial frequency $f_{3}$ = 31.428 day$^{-1}$ ($\ell$=1), and two modulation terms ($f_{m1}$ = 0.065 day$^{-1}$ and $f_{m2}$ = 1.693 day$^{-1}$). In addition, $f_{1}$ is the first overtone mode, which makes KIC 9845907 a $\delta$ Scuti star with the first overtone as the dominant frequency.

Although equidistant frequency-triplet structure has been seen in other $\delta$ Scuti stars, the number of such occurrences in KIC 9845907 is unusual (fourteen pairs; see Figs.~\ref{Triple1} and \ref{Triple2} and Table \ref{tab:T}). The modulation frequencies do not arise from known instrumental effects of \kepler. We also found quintuplet structures in the frequency spectra in KIC 9845907 (see Fig.~\ref{fig:Q} and Table 3). We discussed several potential explanations, i.e., beating, the Blazhko effect, combination mode hypothesis, nonlinear mode coupling, large separation, and stellar rotational splitting for the equidistant structures in frequency spectra. Numerical results of asteroseismic models indicate this modulation with $f_{m1}$ might be related to the rotational splitting. We suggest searching for more $\delta$ Scuti stars with triplet and/or quintuplet structures; using high-precision space photometry would be helpful to explore its origin.

\acknowledgements
 
We would like to thank the \kepler science team for providing such excellent data.
This research is supported by the program of the National Natural Science Foundation of China (grant Nos. U1938104 and 12003020). 
A.G.H. acknowledges support from ``FEDER/Junta de Andaluc\'{\i}a-Consejer\'{\i}a de Econom\'{\i}a y Conocimiento'' under project E-FQM-041-UGR18 by Universidad de Granada and from the Spanish State Research Agency (AEI) project PID2019-107061GB-064.


\section*{Appendix}

In Table~A1, we listed the 85 extracted frequencies (i.e., $f_{1}$ to $f_{85}$), their corresponding amplitudes, and S/N, as well. 

\begin{deluxetable}{ccccccccc} 
\tabletypesize{\small} 
\tablewidth{0pc} 
\tablenum{A1}
\tablecaption{Multi-frequency Solution of the SC Light Curves of KIC 9845907 (denoted by $f_{i}$)
\label{tab:Frequency-SC}} 
\tablehead{ 
\colhead{$f_{i}$}   & 
\colhead{Frequency}  &
\colhead{Amplitude}      &
\colhead{S/N}            &
\colhead{Comment}            \\
\colhead{}   &
\colhead{(day$^{-1}$)}  &
\colhead{(mmag)}      &
\colhead{}            &
\colhead{}            &
}
\startdata 
          & Independent frequencies &  &  &  &\\
        1 & 17.597272(1) & 39.049 & 99.4 & radial &\\ 
        2 & 26.11889(7) & 9.780 & 82.3 & Independent&\\ 
        3 & 31.42813(3) & 8.127 & 56.1 & non-radial &\\ 
        4 & 21.2524(1) & 5.606 & 67.8 & Independent&\\ 
        5 & 29.0141(5) & 3.219 & 51.2 & Independent&\\ 
        6 & 27.0100(6) & 2.435 & 22.8 & Independent&\\ 
        7 & 24.4713(8) & 2.119 & 39.7 & Independent&\\ 
        8 & 0.0649(1) & 1.321 & 28.7 & $f_{m1}$&\\ 
        9 & 1.6927(5) & 0.551 & 20.0 & $f_{m2}$&\\ 
          & Combination frequencies     &       &    &             &\\
        10 & 31.4930(3) & 7.418 & 40.9 & $f_{3}$+$f_{8}$&\\ 
        11 & 31.3631(1) & 4.787 & 26.1 & $f_{3}$$-$$f_{8}$&\\ 
        12 & 21.1456(6) & 2.182 & 23.1 & $f_{2}$+2$f_{5}$$-$$2f_{3}$$-$$2f_{8}$&\\   
        13 & 35.1945(4) & 1.855 & 54.7 & 2$f_{1}$&\\ 
        14 & 23.7585(6) & 1.733 & 31.5 & $f_{2}$+$f_{5}$$-$$f_{3}$$-$$f_{8}$&\\  
        15 & 29.7353(7) & 1.623 & 25.1 & $f_{3}$$-$$f_{9}$&\\  
        16 & 8.5216(2) & 1.280 & 39.6 & $f_{2}$$-$$f_{1}$&\\ 
        17 & 62.8562(6) & 1.248 & 58.3 & 2$f_{3}$&\\ 
        18 & 59.4707(5) & 1.241 & 54.5 & 2$f_{3}$$-$$2f_{9}$&\\ 
        19 & 61.1635(1) & 1.166 & 53.2 & 2$f_{3}$$-$$f_{9}$&\\ 
        20 & 28.9573(1) & 0.845 & 13.6 & $f_{5}$$-$$f_{8}$&\\ 
        21 & 13.8308(6) & 0.799 & 18.9 & $f_{3}$$-$$f_{1}$&\\ 
        22 & 13.8957(6) & 0.785 & 18.2 & $f_{3}$$-$$f_{1}$+$f_{8}$&\\  
        23 & 49.0254(5) & 0.582 & 27.9 & $f_{1}$+$f_{3}$&\\ 
        24 & 26.8801(3) & 0.581 & 23.8 & $f_{6}$$-$$f_{8}$&\\ 
        25 & 77.0680(3) & 0.571 & 45.8 & $f_{1}$+2$f_{3}$$-$$2f_{9}$&\\       
        26 & 47.3326(5) & 0.553 & 29.0 & $f_{1}$+$f_{3}$$-$$f_{9}$&\\ 
        27 & 43.7161(6) & 0.473 & 24.6 & $f_{1}$+$f_{2}$&\\ 
        28 & 13.7659(5) & 0.472 & 12.1 & $f_{3}$$-$$f_{1}$$-$$f_{8}$&\\ 
        29 & 49.0903(1) & 0.459 & 21.5 & $f_{1}$+$f_{3}$+$f_{8}$&\\ 
        30 & 24.3602(1) & 0.439 & 8.3 & $f_{7}$$-$$2f_{8}$&\\ 
        31 & 52.2379(9) & 0.416 & 21.6 & 2$f_{2}$&\\ 
        32 & 29.6703(6) & 0.403 & 6.3 & $f_{3}$+$f_{7}$$-$$f_{2}$$-$$2f_{8}$&\\ 
        33 & 5.3741(4) & 0.392 & 12.6 & $f_{3}$+$f_{8}$$-$$f_{2}$&\\ 
        34 & 5.3091(2) & 0.391 & 12.7 & $f_{3}$$-$$f_{2}$&\\ 
        35 & 61.2284(1) & 0.389 & 18.5 & 2$f_{3}$+$f_{8}$$-$$f_{9}$&\\ 
        36 & 29.8002(8) & 0.379 & 8.2 & $f_{3}$+$f_{7}$$-$$f_{2}$&\\ 
        37 & 10.1756(1) & 0.376 & 13.6 & $f_{3}$$-$$f_{4}$&\\ 
        38 & 48.9605(1) & 0.374 & 17.5 & $f_{1}$+$f_{3}$$-$$f_{8}$&\\ 
        39 & 24.4160(2) & 0.364 & 6.8 & $f_{2}$$-$$f_{9}$&\\ 
        40 & 78.7607(8) & 0.359 & 31.7 & $f_{1}$+2$f_{3}$$-$$f_{9}$&\\ 
        41 & 38.8497(8) & 0.348 & 17.6 & $f_{1}$+$f_{4}$&\\ 
        42 & 11.4168(7) & 0.335 & 12.3 & $f_{5}$$-$$f_{1}$&\\ 
        43 & 54.1515(2) & 0.308 & 20.4 & $f_{3}$+2$f_{7}$$-$$f_{2}$$-$$2f_{8}$&\\ 
        44 & 27.8969(3) & 0.299 & 9.6 & $f_{1}$+3$f_{3}$+2$f_{8}$$-$$f_{2}$$-$$2f_{5}$&\\ 
        45 & 44.6073(3) & 0.297 & 17.7 & $f_{1}$+$f_{6}$&\\ 
        46 & 57.5470(3) & 0.296 & 16.8 & $f_{2}$+$f_{3}$&\\ 
        47 & 80.4535(8) & 0.293 & 28.7 & $f_{1}$+2$f_{3}$&\\ 
        48 & 62.9211(7) & 0.288 & 13.0 & 2$f_{3}$+$f_{8}$&\\ 
        49 & 13.9535(1) & 0.287 & 7.5 & 2$f_{1}$$-$$f_{4}$&\\ 
        50 & 9.41278(8) & 0.276 & 10.1 & $f_{6}$$-$$f_{1}$&\\ 
        51 & 4.4830(9) & 0.275 & 9.9 & $f_{3}$+$f_{8}$$-$$f_{6}$&\\ 
        52 & 90.8987(7) & 0.266 & 32.0 & 3$f_{3}$$-$$2f_{9}$&\\ 
        53 & 57.6120(4) & 0.262 & 15.2 & $f_{2}$+$f_{3}$+$f_{8}$&\\ 
        54 & 61.0984(9) & 0.261 & 12.2 & 2$f_{3}$$-$$f_{8}$$-$$f_{9}$&\\ 
        55 & 54.8187(5) & 0.258 & 15.9 & $f_{2}$+$f_{6}$+$f_{9}$&\\ 
        56 & 58.3731(7) & 0.244 & 16.5 & $f_{3}$+$f_{6}$$-$$f_{8}$&\\ 
        57 & 3.6551(2) & 0.241 & 9.9 & $f_{4}$$-$$f_{1}$&\\ 
        58 & 5.2443(3) & 0.238 & 7.5 & $f_{3}$$-$$f_{2}$$-$$f_{8}$&\\ 
        59 & 10.2405(2) & 0.236 & 7.8 & $f_{3}$+$f_{8}$$-$$f_{4}$&\\ 
        60 & 55.1330(4) & 0.228 & 13.7 & $f_{2}$+$f_{5}$&\\ 
        61 & 41.8733(7) & 0.219 & 11.5 & 2$f_{3}$$-$$f_{1}$$-$$2f_{9}$&\\ 
        62 & 52.6805(3) & 0.213 & 10.8 & $f_{3}$+$f_{4}$&\\ 
        63 & 3.2188(6) & 0.212 & 7.5 & $f_{7}$$-$$f_{4}$&\\ 
        64 & 62.7913(6) & 0.220 & 10.0 & 2$f_{3}$$-$$f_{8}$&\\ 
        65 & 1.7576(5) & 0.209 & 7.4 & $f_{2}$+2$f_{8}$$-$$f_{7}$&\\ 
        66 & 55.6641(9) & 0.198 & 12.5 & 2$f_{1}$+4$f_{3}$+2$f_{8}$$-$$f_{2}$$-$$f_{4}$$-$$2f_{5}$&\\ 
        67 & 42.3981(8) & 0.189 & 9.5 & $f_{2}$+$f_{4}$+2$f_{5}$$-$$2f_{3}$$-$$2f_{8}$&\\ 
        68 & 2.4788(8) & 0.183 & 6.5 & $f_{3}$+$f_{8}$$-$$f_{5}$&\\ 
        69 & 7.6695(6) & 0.182 & 7.3 & 2$f_{3}$+$f_{8}$$-$$f_{2}$$-$$f_{5}$&\\ 
        70 & 55.8442(7) & 0.177 & 9.9 & $f_{3}$+$f_{7}$$-$$f_{8}$&\\ 
        71 & 42.5049(1) & 0.176 & 10.1 & 2$f_{4}$&\\ 
        72 & 57.4821(2) & 0.176 & 8.2 & $f_{2}$+$f_{3}$$-$$f_{8}$&\\ 
        73 & 55.8995(1) & 0.175 & 9.8 & $f_{3}$+$f_{7}$&\\ 
        74 & 6.9567(4) & 0.169 & 6.8 & $f_{3}$$-$$f_{7}$&\\ 
        75 & 45.2589(9) & 0.167 & 9.4 & 2$f_{3}$$-$$f_{1}$&\\ 
        76 & 75.1444(1) & 0.128 & 14.2 & $f_{1}$+$f_{2}$+$f_{3}$ &\\
        77 & 78.8256(8) & 0.126 & 11.3 & $f_{1}$+2$f_{3}$+$f_{8}$$-$$f_{9}$ &\\ 
        78 & 78.6958(8) & 0.123 & 10.6 & $f_{1}$+2$f_{3}$$-$$f_{8}$$-$$f_{9}$&\\ 
        79 & 75.2092(7) & 0.096 & 10.8 & $f_{1}$+$f_{2}$+$f_{3}$+$f_{8}$&\\
        80 & 90.9639(1) & 0.085 & 9.8  & 3$f_{3}$+$f_{8}$$-$$2f_{9}$&\\ 
        81 & 75.0793(9) & 0.075 & 8.0  & $f_{1}$+$f_{2}$+$f_{3}$$-$$f_{8}$&\\
        82 & 94.2843(9) & 0.069 & 9.8  & 3$f_{3}$ &\\
        83 & 94.3493(4) & 0.069 & 9.7  & 3$f_{3}$+$f_{8}$ &\\
        84 & 90.8339(8) & 0.062 & 7.3  & 3$f_{3}$$-$$f_{8}$$-$$2f_{9}$&\\ 
        85 & 94.2194(9) & 0.048 & 6.9  & 3$f_{3}$$-$$f_{8}$ &\\
    
   \enddata 

\end{deluxetable}

\end{CJK}
\end{document}